\documentclass[aps,pre,reprint,longbibliography,amsmath,amsfonts,amssymb,showpacs]{revtex4-1}
\usepackage{graphicx,CJK}
\begin{document}
\begin{CJK*}{UTF8}{mj}
\title{Universality-class crossover by a nonorder field introduced to the pair contact process with diffusion }
\author{Su-Chan Park (박수찬)}
\affiliation{Department of Physics, The Catholic University of Korea, Bucheon 14662, Republic of Korea}
\date{\today}
\begin{abstract}
The one-dimensional pair contact process with diffusion (PCPD), an interacting particle 
system with diffusion, pair annihilation, and creation by pairs, has defied a consensus about the
universality class that it belongs to.
An argument by Hinrichsen [H. Hinrichsen, Physica A {\bf 361}, 457 (2006)] 
claims that freely diffusing particles in the PCPD should play the same role as
frozen particles, when it comes to the critical behavior.
Therefore, the PCPD is claimed to have the same critical phenomena as 
a model with infinitely many absorbing states that belongs to the directed percolation (DP) universality
class.  To investigate if diffusing particles are really indistinguishable from 
frozen particles in the sense of the renormalization group, we numerically study a variation of the PCPD by 
introducing a nonorder field associated with infinitely
many absorbing states. We find that a crossover from the PCPD to the DP occurs
due to the nonorder field. Since, by studying a similar model, we exclude the possibility
that mere introduction of a nonorder field to one model 
can entail a nontrivial crossover to another model in the same universality class, we attribute
the observed crossover to the difference of the universality class of the PCPD from the DP class.
\end{abstract}
\maketitle
\end{CJK*}
\section{Introduction}
The pair contact process with diffusion (PCPD) is an interacting particle system
with diffusion, pair annihilation ($2A \rightarrow 0$), and creation by pairs ($2A \rightarrow 3A$),
which exhibits an absorbing phase transition.
The PCPD was first studied by Grassberger in 1982~\cite{G1982}, 
but it had not attracted much interest until Howard and T\"auber~\cite{HT1997}
introduced a so-called `bosonic' PCPD. 
Since $2A \rightarrow 3A$ dynamics in the bosonic PCPD drives particle density in the active phase to 
infinity in finite time, a more controllable model was desired to study the phase transition. 
Carlon et al.~\cite{CHS2001} introduced a present form of the PCPD on a lattice with hard core exclusion
to keep the density from diverging.
The conclusion in Ref.~\cite{CHS2001} was already
controversial, which has triggered a lot of numerical and analytical studies of the PCPD~\cite{H2001,O2000,PHK2001,H2001_291,PK2002,DdM2002,KC2003,O2003,BC2003,JvWDT2004,PHP2005a,PHP2005b,H2006,PP2006,KK2007,SB2008,PP2009,SB2012,GCDD2014,SCP2014Cde}. 

It turned out that the PCPD, especially in one dimension, is influenced by strong corrections to scaling.
This, in turn, makes it difficult for numerical studies to lead to a consensus about the universality class 
of the one dimensional PCPD. Accordingly, many scenarios had been suggested 
in the early stage of research (for a review of early discussions, see~\cite{HH2004,PP2008EPJB}). 
The controversy remains unabated, but by now only two competing theories have survived.

In one theory, the PCPD is claimed to form a different universality class from the directed percolation (DP)
universality class~\cite{KC2003,PHP2005a,PHP2005b,PP2006,PP2009,SCP2014Cde}. 
This theory is supported by the following facts:
First, the upper critical dimension of the PCPD is 2~\cite{OMS2002}, while that of the DP is 4.
Second, diffusion bias changes the universality class of the one-dimensional PCPD, whereas
such a bias can be asymptotically removed by a Galilean transformation in the DP class~\cite{PHP2005a}.
Third, there are nontrivial crossover behaviors between the PCPD and the DP models~\cite{PP2006,PP2009}. 
In the other theory, the PCPD is claimed to belong to
the DP class~\cite{BC2003,H2006,SB2008,SB2012}.
An argument to support this theory was put forward by Hinrichsen~\cite{H2006},
which will be called {\em the DP argument} for later reference.

Since the DP argument has motivated the present work, we repeat it here for completeness.
The DP argument is based on a numerical observation that the dynamic exponent $z$ of the PCPD
in one dimension is smaller than 2. Since critical clusters spread as $t^{1/z}$ and isolated particles
spread diffusively as $t^{1/2}$, a comparison of these two scales 
suggests that the critical spreading will eventually dominate the critical behavior over 
the diffusive spreading 
and, accordingly, diffusion is irrelevant in the sense of the renormalization group (RG).
Although it is in principle possible for a pair to be formed purely by diffusion of isolated particles, 
the irrelevance of diffusion implies that such events will hardly occur in the long time limit. 
Thus, it does not matter whether a particle can diffuse or not, as far as critical phenomena are concerned.
If particles are not allowed to diffuse, the model is the pair contact process 
(PCP)~\cite{J1993}, which has infinitely many absorbing states (IMAS) and belongs to the DP class.
In this sense, diffusion can at best affect corrections to scaling 
and the one-dimensional PCPD should exhibit the same universal behavior as the PCP or, 
in the context of universality, as the DP.

The aim of this paper is to figure out if diffusing particles in the PCPD are really indistinguishable in the RG sense from the frozen isolated particles of the PCP, as claimed in the DP argument.
If isolated particles can be regarded as frozen particles of the PCP, introducing
dynamics which make diffusing particles frozen should not yield any other singular behavior than
the typical DP critical behavior. Motivated by this idea, we introduce and study numerically
a variation of the PCPD in which 
diffusing particles can mutate irreversibly to immobile species with rate $w$.
The case with $w=0$ will correspond to the PCPD and  
the case with nonzero $w$, which includes the PCP ($w=1$), has IMAS. The exact definition of
the model will be given in Sec.~\ref{Sec:model}.
We will present simulation results of the model in Sec.~\ref{Sec:results}, focusing on the existence
of a crossover around $w=0$.
Section~\ref{Sec:sum} summarizes and concludes this work.

\section{\label{Sec:model}Model}
This section introduces a one-dimensional two-species lattice model 
with periodic boundary conditions.
The system size will be denoted by $L$, and each species will
be denoted by $A$ and $B$, respectively.
Each site can be one of the three possible states; $A$-occupied, $B$-occupied, and vacant states.
No multiple occupancy is allowed.
The system evolves stochastically under the following rules:
\begin{align}
\label{Eq:rules}
&\varnothing A \stackrel{w/2}{\longrightarrow} \varnothing B,\quad
A \varnothing \stackrel{w/2}{\longrightarrow} B\varnothing,\quad
\varnothing A \stackrel{D}{\longleftrightarrow}A \varnothing,\\
&XY \stackrel{p}{\longrightarrow} \varnothing \varnothing,\quad
XY\varnothing \stackrel{\sigma}{\longrightarrow} XYA,\quad
\varnothing XY \stackrel{\sigma}{\longrightarrow} AXY,
\nonumber 
\end{align}
where $\varnothing$ stands for a vacant site; $A$ ($B$) represents a site occupied
by species $A$ ($B$); $X$ and $Y$ can be any of $A$ and $B$; and the variables
above the arrows mean the transition rates of the corresponding events. 
In all simulations in this paper, we set $\sigma = (1-p)/2$ and $D=(1-w)/2$
with two parameters $w$ and $p$ ($0\le w \le 1$ and $0 \le p \le 1$). 
We always use the configuration with all sites occupied by species $A$ as an 
initial condition and all discussions are based on this initial condition.
The extension to higher dimensions is straightforward, but most discussions in this paper
are limited to the one dimensional model.

For the case with $w=0$, species $B$ cannot be generated by the rules, so this
case is identical to the PCPD studied in Ref.~\cite{PHP2005a}. 
When $w$ is strictly positive, $B$ particles can appear by the dynamics and
any configuration only with isolated $B$ particles are absorbing.
Hence, the model with nonzero $w$ has infinitely many absorbing states, which is characterized
by the nonzero density of $B$ species in all phases.
Following the convention, we will call the density of $B$ species, denoted by $\rho_b$, a nonorder field.
In particular, if $w=1$, isolated $A$ particles can at best mutate to $B$ particles, so
the two species play the exactly same role as particles in the PCP~\cite{J1993}.
Since our main concern is the regime where $0\le w \ll 1$,
we will call this model the PCPD with a nonorder field (PNF).

To simulate the model, we made a list of active pairs. By an active pair we mean
two consecutive sites that take one of 6 forms, $\varnothing A$, $A\varnothing $, 
$AA$, $AB$, $BA$, and $BB$. The list contains the information as to where active pairs are 
located. The size of the list at time $t$ is denoted by $N_t$.
At $t=0$ when all sites are occupied by $A$, the size of the list is the same as the system 
size $L$. At time $t$, we choose one of the active pairs at random with equal probability. 
If only one site is occupied (by $A$ by definition) in the chosen active pair, this particle mutates 
to $B$ with probability $w/2$ or hops to the empty site inside the chosen active pair with probability 
$D = (1-w)/2$. With probability $\frac{1}{2}$, however, nothing happens. If 
both sites of the chosen active pair are occupied,
with probability $p$ these two sites become vacant ($XY \rightarrow
\varnothing \varnothing$) or with probability $1-p$ there will be an attempt to branch an $A$ particle to one of its 
nearest neighbors, to be called a target site, which is chosen at random with equal probability. 
If the target site is empty, 
$A$ particle will be placed there. Otherwise, nothing happens. After the above step, 
time increases by $1/N_t$. The above procedure will continue until 
either the system falls into one of the absorbing states or time exceeds 
the preassigned observation time.
In all simulations whose results will be presented in the next section, however,
the system size is so large that no simulation ends up with an absorbing state at the
end of the preassigned observation time.

We are mainly interested in the behavior of pair density defined as
\begin{equation}
\rho_p(t) \equiv \frac{1}{L} \sum_{i=1}^L \left \langle 
\left (1-\delta_{X_i(t),\varnothing}\right ) 
\left (1-\delta_{X_{i+1}(t),\varnothing}\right )\right \rangle,
\end{equation}
where $X_i(t)$ means the state (one of $A$, $B$, and $\varnothing$) 
at site $i$ at time $t$,
$\delta$ is the Kronecker delta symbol, and $\langle \ldots \rangle$ stands for
the average over all realizations.
The order parameter is then defined as
\begin{align}
\rho_s = \lim_{t\rightarrow \infty} \lim_{L\rightarrow \infty} \rho_p(t),
\label{Eq:order_param}
\end{align}
which is zero (nonzero) in the absorbing (active) phase.
Note that $\rho_s$ defined in Eq.~\eqref{Eq:order_param} is the order parameter
used in both the PCP and the PCPD.

Before presenting simulation data, we would like to argue that the PNF with nonzero
$w$ should belong to the DP class, mimicking the DP argument with a slight modification. 
As long as $w$ is nonzero, an isolated $A$ particle can mutate to $B$ within finite time of $O(1/w)$
for any $p$. Since the correlation time diverges at the critical point, 
diffusion of $A$ particles which should be terminated in finite time is irrelevant 
in the RG sense.
Thus, the critical behavior of the PNF with nonzero $w$ cannot be distinguishable from the PCP which
belongs to the DP class.

Note that, in the original DP argument, the dynamic exponent has to be assumed to be smaller than 2, which
makes the extension to higher dimensional systems problematic.
In our argument, however, we do not have to assume the value of the dynamic
exponent, so it can also be applicable to higher dimensions.

\section{\label{Sec:results}Results}
This section presents simulation results of the PNF defined by the rules in Eq.~\eqref{Eq:rules}. 
We use $p$ as a tuning parameter with $w$ fixed and
the critical point will be denoted by $p_c(w)$.
The system size is $L=2^{20}$ and
no realization has ended up with an absorbing state within the observation time,
which minimally guarantees that the finite size effect is not important.

\begin{figure}
\includegraphics[width=\columnwidth]{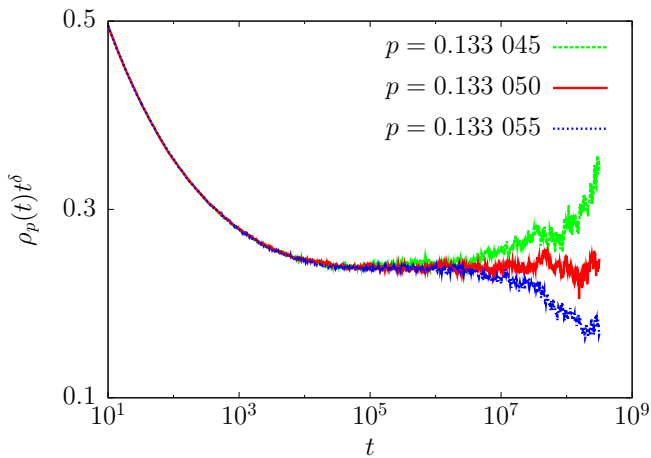}
\caption{\label{Fig:DP1e4} (Color online) Semilogarithmic plots of $\rho_p(t) t^\delta$ vs. $t$ around the
critical point for $w=10^{-4}$, where $\delta = 0.1595$ is the critical decay exponent of the DP.
For $p=0.133~05$ (middle curve), the curve becomes flat from $t=10^5$ for about
three log decades and the other curves veer up ($p=0.133~045$) or down ($p=0.133~055$). 
Thus, we conclude that the critical point is $p_c=0.133~050(5)$ and
the model belongs to the DP class.
}
\end{figure}
\begin{table}[b]
\caption{\label{Table:pc} Critical points of the PNF for various $w$'s. The numbers
in parentheses indicate the errors of the last digits.}
\begin{ruledtabular}
\begin{tabular}{llll}
$w$&$p_c(w)$&$w$&$p_c(w)$\\
\hline
0&$0.133~519(3)$\footnotemark[1]&0.1&0.103~635(5)  \\
0.0001&0.133~050(5)&0.2&0.093~545(5)  \\
0.0003&0.132~595(5)&0.3&0.087~872(2)  \\
0.0008&0.131~83(1) &0.4&0.084~265(5)  \\
0.001&0.131~580(5)&0.5&0.081~815(5)  \\
0.002&0.130~54(1) &0.6&0.080~090(5)  \\
0.004&0.128~92(1) &0.7&0.078~860(5)  \\
0.008&0.126~45(1) &0.8&0.077~990(5)  \\
0.01   &0.125~410(5)&0.9&0.077~410(5)  \\
0.05   &0.112~56(1) &1&$0.077~0905(5)$\footnotemark[2] 
\end{tabular}
\end{ruledtabular}
\footnotetext[1]{From Ref.~\cite{PHP2005a}.}
\footnotetext[2]{From Ref.~\cite{PP2007}.}
\end{table}
We first show the behavior of $\rho_p(t)$ around the critical point for $w=10^{-4}$.
In Fig.~\ref{Fig:DP1e4}, we plot $\rho_p(t) t^{\delta}$ against $t$ around the critical point, 
where $\delta = 0.1595$ is the critical decay exponent of the DP class. 
In what follows, the symbol $\delta$ is reserved for the DP critical decay exponent.
The numbers of independent runs are 16, 24, and 24 for $p=0.133~045$, $0.133~05$, and $0.133~055$, respectively.
At $p=0.133~05$, the curve is almost flat for more than three log decades, while two
other curves veer up or down in the long time limit. 
Hence, we conclude that this case indeed belongs to the DP class with $p_c = 0.133~050(5)$,
where the number in parentheses is the uncertainty of the last digit.
We also studied the critical behavior for various values of $w$ in a similar manner.
The resulting critical points are summarized in Table~\ref{Table:pc}.

\begin{figure}
\includegraphics[width=\columnwidth]{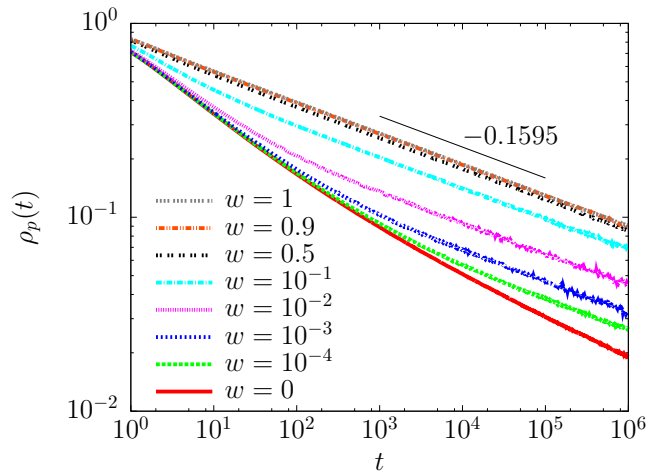}
\caption{\label{Fig:rp} (Color online) Double-logarithmic plots of $\rho_p$ vs. $t$ at the critical points
for various $w$'s from $w=0$ to $w=1$ (bottom to top).
Three curves corresponding
to $w=0.5, 0.9$ and 1 are barely discernible, whereas difference among curves
corresponding to $w<0.1$ are conspicuous.
}
\end{figure}

Since both the PCP and the PNF with $0 < w < 1$ belong to the DP universality class, 
there should not be a significant change due to the variation of $w$ when $(1-w) \ll 1$.
To make this point strong, we present the behaviors of $\rho_p(t)$ at the corresponding
critical points to various values of $w$ in Fig.~\ref{Fig:rp}.
Indeed, the critical decays of $\rho_p(t)$ for $w \ge 0.9$ are barely discernible.

A symptom of crossover is already observed in Fig.~\ref{Fig:rp} for $w \ll 1$.
When $w < 0.1$, $\rho_p(t)$'s at $p=p_c(w)$ transiently follow the PCPD behavior and then eventually deviate from the PCPD curve.
In particular, the time when the critical PNF starts to deviate from the PCPD curve increases as $w$ gets smaller and
there is no symptom of saturation up to $w = 10^{-4}$. 
This behavior seems consistent with the crossover scaling ansatz~\cite{PP2006}
\begin{align}
\rho_p(t;\Delta,w) = t^{-\alpha} F_c\left ( \Delta w^{-1/\phi},  t w^{\nu_\|/\phi}\right ),
\end{align}
where $\Delta \equiv p_0 - p$ with $p_0$ to be the critical point of the model with $w=0$ (that is, the critical point of the PCPD), $\alpha$
is the critical decay exponent of the PCPD, $\nu_\|$ is the correlation time exponent of the PCPD,
$\phi$ is the crossover exponent, and $F_c$ is a scaling function.

Since the scaling ansatz suggests a power-law behavior of the phase boundary for small $w$ in such a manner that 
\begin{equation}
\label{Eq:cross}
p_0 - p_c(w) \sim w^{1/\phi},
\end{equation}
we can estimate $\phi$ by analyzing the behavior of the phase boundary for 
small $w$ without resorting to the values of $\alpha$ and $\nu_\|$.
Although $\alpha$ was estimated as $\approx 0.19$ in Ref.~\cite{PHP2005a},
which is larger than the more accurate estimate in Ref.~\cite{SCP2014Cde}, 
one should note that the accuracy of $p_0$ in Ref.~\cite{PHP2005a} was
attained without resorting much to the accuracy of $\alpha$; the critical
point was actually estimated by finding two $p$'s belonging to the absorbing and active phases, respectively. 
Thus, we can safely use the value $p_0$ from Ref.~\cite{PHP2005a} to find the crossover exponent $\phi$.
\begin{figure}
\includegraphics[width=\columnwidth]{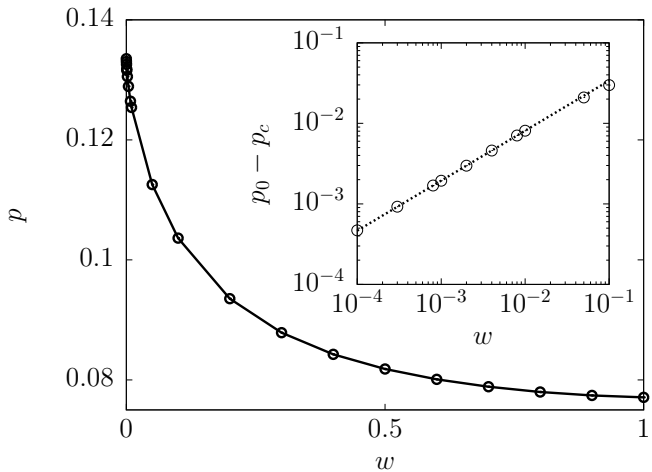}
\caption{\label{Fig:phi} Phase boundary of the PNF in the $wp$-plane. 
Symbols represent the critical points obtained by simulations (see Table~\ref{Table:pc}) and lines which connect two nearest symbols
are for guides to the eyes.
The phase boundary approaches the ordinate continuously with infinite slope, while it approaches the 
PCP critical point (corresponding to $w=1$) with
finite slope. Inset: Plot of $p_0 - p_c$ vs. $w$ on a double logarithmic scale. For
comparison, Eq.~\eqref{Eq:cross} with $1/\phi \approx 0.62$ is also drawn. 
}
\end{figure}

In Fig.~\ref{Fig:phi}, we depict the phase boundary of the PNF in the $wp$-plane. 
The phase boundary approaches the ordinate with infinite slope,
which suggests that $1/\phi$ in Eq.~\eqref{Eq:cross}
is smaller than 1. The fitting of the phase boundary gives $1/\phi = 0.62(2)$; see the inset
of Fig.~\ref{Fig:phi}. Hence, we conclude that there is indeed a crossover behavior 
for $w\ll 1$ and the time for the critical PNF to deviate from the critical decay of the PCPD
diverges as $w$ gets smaller.

Although the number of absorbing states in itself does not depend on $w$ for any $L$, 
the so-called natural density~\cite{JD1993} of the nonorder field $\rho_b$ 
is expected to vary with $w$. Since the discontinuity of the natural density 
of a nonorder field can give a crossover even within the same universality
class~\cite{PP2007}, we need to check if the natural density changes abruptly at $w=0$. 

\begin{figure}
\includegraphics[width=\columnwidth]{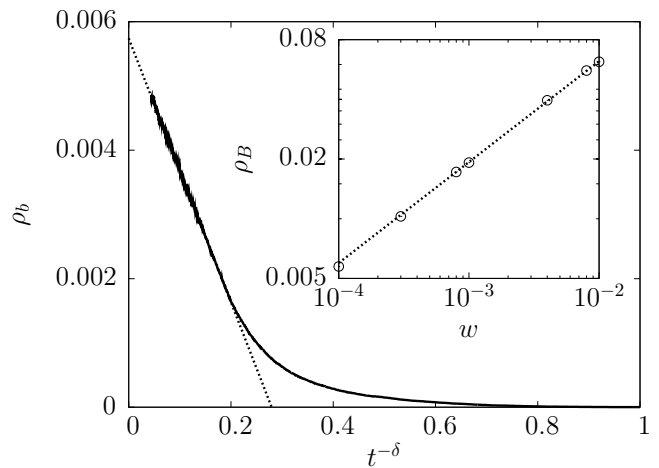}
\caption{\label{Fig:rhoB} Plot of $\rho_b$ vs. $t^{-\delta}$ at the
critical point $p_c(w)$ for $w=10^{-4}$. Close to the ordinate, the curve becomes almost straight.
By a linear extrapolation using Eq.~\eqref{Eq:rhob}, we found the natural density $\rho_B$. 
The fitting result is drawn as a straight line.
Inset: Plot of $\rho_B$ vs. $w$ on a double logarithmic scale. $\rho_B$ for small $w$ is fitted
by a power function $a w^\chi$ with a result $\chi \approx 0.51$.
}
\end{figure}

At the critical point, the density $\rho_b(t)$ of $B$ species is expected to have
asymptotic behavior as~\cite{OMSM1998}
\begin{align}
\rho_b(t) -\rho_B(w)  \sim t^{-\delta},
\label{Eq:rhob}
\end{align}
where $\rho_B(w)$ is the natural density at the critical point for given $w$. 
If we plot $\rho_b(t)$ as a function of $t^{-\delta}$,
a straight line is expected and the extrapolation of the straight line to $t\rightarrow \infty$
can give $\rho_B(w)$. In Fig~\ref{Fig:rhoB}, we show the result
of this extrapolation for $w= 10^{-4}$. Applying the same method,
we obtained $\rho_B$ for other values of $w$. In the inset of Fig.~\ref{Fig:rhoB},
we show how $\rho_B$ behaves with $w$. As can be easily guessed,
$\rho_B$ is an increasing function of $w$. Furthermore, it turns out that $\rho_B$ 
exhibits a power-law behavior as $\sim w^{\chi}$ with an estimate $\chi \approx 0.51$, which
suggests that $\rho_B$ is continuous at $w=0$.

We now argue that $\chi$ is exactly 0.5.
First, we take the existence of nonzero $\rho_B(w)$ for granted.
Now, consider a situation where an isolated $A$ particle appears (by the given dynamics) in the sea of $B$ particles.
This isolated $A$ particle will either mutate to $B$ in time of $O(1/w)$ or meet another $B$ 
particle in time of $O(1/\rho_B^2)$ ($\sqrt{Dt} \sim 1/\rho_B$ with $1/\rho_B$ to be the order of mean distance between $A$ and the nearest $B$ particle in this situation). If $\rho_B \ll \sqrt{w}$, 
the chance of meeting a $B$ particle before mutation occurs is very low. 
Thus, we can regard isolated $A$ particles as frozen $B$ particles. 
In this case, $\rho_B$ is purely determined
by the dynamics of pairs and should not depend on $w$, which contradicts to the assumption 
$\rho_B \ll \sqrt{w}$.
If $\rho_B \gg \sqrt{w}$, diffusing $A$ particle will find a frozen $B$ particle before it mutates and
the newly formed pair begins so-called defect dynamics. 
Since the system is at the critical point, the defect dynamics are eventually terminated and
the number of $B$ particles is likely to decrease by 1 at the end of the defect dynamics
when it is compared with the number of $B$ particles at the time a pair is formed by diffusion.
As the number of $B$ particles can only decrease as long as $\rho_B$ is finite,
$\rho_B$ should decrease to zero, which again is contradictory to the assumption $\rho_B \gg \sqrt{w}$.
Thus, a consistent conclusion is attainable only for $\rho_B \sim \sqrt{w}$.

Finally, we would like to check if mere presence of a nonorder field 
is enough to result in a crossover even within the same universality class.
To this end, we slightly modify the PCP (that is, the PNF with $w=1$) in such a way that
whenever isolated particles appear by a pair annihilation,  
each isolated particle is removed with probability $1-\varepsilon$ and
it remains there with probability $\varepsilon$. 
Note that up to two isolated particles can appear
by a single pair-annihilation event (if two $A$ particles
in the middle of a local configuration $\varnothing AAAA \varnothing$ are removed by pair annihilation, 
two $A$'s at the end of this cluster become isolated).
Once an isolated particle which appears right after a pair-annihilation event is decided not to disappear, 
it remains frozen until another particle come to one of its nearest neighbors.
We use the same transition rates as in Eq.~\eqref{Eq:rules} for creation and annihilation events.
For convenience, we will call this model the modified PCP (MPCP).

The MPCP with $\varepsilon=1$ is just the PCP. When $\varepsilon=0$, there is no isolated particles and this 
becomes a model with a single absorbing state (we take a configuration with every site occupied by a particle
as an initial state).  Just like the PNF, the number of absorbing states does not depend 
on $\varepsilon$, though the natural density of a nonorder field 
should depend on $\varepsilon$. For any value of $\varepsilon$, MPCP is expected to belong to the DP 
universality class. 

\begin{figure}
\includegraphics[width=\columnwidth]{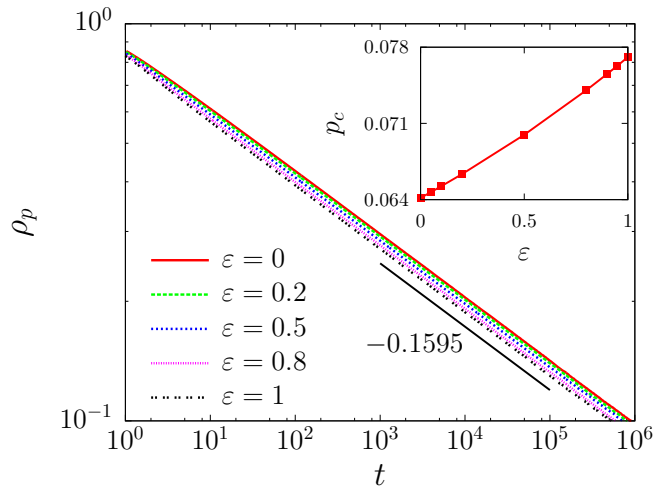}
\caption{\label{Fig:pc_dp} (Color online) Double-logarithmic plots of $\rho_p$ vs. $t$ at the 
critical point of the MPCP for various $\varepsilon$'s from $\varepsilon=0$ to $\varepsilon=1$ (top to bottom). As a guide to the eyes, a line segment with slope $-0.1595$ is
also drawn. Unlike the PNF, the critical behavior for small $\varepsilon$ is hardly
discernible from that for $\varepsilon=0$; see also Fig.~\ref{Fig:rp}.
Inset: Plot of $p_c$ as a function of $\varepsilon$ for the MPCP. Unlike the PNF, the slope at 
$\varepsilon = 0$ is finite.
}
\end{figure}

Using the DP exponent $\delta$, we estimated critical points by studying the behavior
of $\rho_p t^\delta$ as in Fig.~\ref{Fig:DP1e4}.
The critical points are summarized in Table~\ref{Table:pc_D}.
To see whether there is a nontrivial crossover for small $\varepsilon$, we plotted
critical decay behavior for various $\varepsilon$'s in Fig.~\ref{Fig:pc_dp}. 
Unlike the PNF around $w=0$, there is no symptom of a diverging time scale for $\varepsilon \ll 1$.
Furthermore, $p_c$ of the MPCP has finite slope at $\varepsilon = 0$ unlike
the PNF at $w=0$.
Thus, we conclude that mere introduction of a nonorder field is not sufficient to trigger a crossover.

\begin{table}[t]
\caption{\label{Table:pc_D} Critical points of the MPCP for various $\varepsilon$'s. The numbers
in parentheses indicate the errors of the last digits.}
\begin{ruledtabular}
\begin{tabular}{llll}
$\varepsilon$&$p_c$&$\varepsilon$&$p_c$\\
\hline
0   &0.064~162(1)  &0.8 &0.074~046(1)    \\
0.05&0.064~6925(5) &0.9 &0.075~5325(15)  \\
0.1 &0.065~234(1)  &0.95&0.076~303(1)    \\
0.2 &0.066~3465(5) &1   &0.077~0905(5)$\footnotemark[1]$  \\
0.5 &0.069~957(1)  &&
\end{tabular}
\end{ruledtabular}
\footnotetext[1]{From Ref.~\cite{PP2007}.}
\end{table}

%
\section{\label{Sec:sum}Summary and discussion}
We studied a variation of the pair contact process with diffusion (PCPD) 
by introducing a nonorder field assocated with infinitely many absorbing states; see Eq.~\eqref{Eq:rules}. 
We called this model the PCPD with a nonorder field (PNF). 
We analyzed absorbing phase transitions for various values of $w$. 
When $w$ is nonzero, the PNF has infinitely many absorbing states (IMAS) and was found to belong
to the directed percolation (DP) universality class. 
When $w$ is close to 1, critical decay of the pair density $\rho_p(t)$ does not show any significant
change from the pair contact process (PCP). 
This is also manifest by the finite slope of the phase boundary at $w=1$; see Fig.~\ref{Fig:phi} around $w=1$.
On the other hand, a singular behavior representing a crossover appears when nonorder field is introduced
to the PCPD (the PNF with $w=0$). This crossover is described by the crossover exponent $\phi$,
whose numerical value was found as $1/\phi \approx 0.62$ from the analysis of the phase boundary at $w=0$ 
(see Fig.~\ref{Fig:phi} around $w=0$ and its inset).
We argued that the natural density $\rho_B$ at the critical point changes continuously as $\sqrt{w}$ 
when $w \ll 1$, which is consistent with simulation results.
To support that this crossover is originated from the difference of the universality classes,
we also presented simulation results of another model which belongs to the DP class irrespective of
whether the number of absorbing states is infinite or finite.
Unlike the PNF around $w=0$, no crossover was observed in this case.

In Refs.~\cite{PP2006,PP2009}, the existence of the nontrivial crossover from the PCPD to the DP 
was invoked to be the evidence of the existence of the fixed point of the PCPD distinct from
the DP in one dimension. The same conclusion is also arrived at in this paper.
Furthermore, our work clarifies the difference between diffusing particles in the PCPD
and frozen particles in the PCP, unlike the anticipation from the DP argument.

It is worth while to discuss the role of a nonorder field which sometimes triggers
a crossover within the same universality class~\cite{PP2007} and sometimes not.
This difference can be understood as follow:
The nonorder field in this paper cannot play a role of an order parameter for all cases, whereas
the nonorder field in Ref.~\cite{PP2007} becomes an order parameter as soon as
crossover dynamics are introduced. That is, the crossover dynamics in Ref.~\cite{PP2007} make
the irrelevant nonorder field a relevant order parameter, which is the origin of 
the crossover within the same universality class. 
observed in Ref.~\cite{PP2007} reflects that a nonorder field becomes relevant in the RG sense
by crossover dynamics while absence of role change of nonorder field in this paper does not
give a nontrivial crossover behavior within the same universality class.

As a final remark, we would like to compare the three different crossovers in this work and  in Ref.~\cite{PP2006,PP2009}
to the also three different crossovers from the directed Ising (DI) class to the DP class~\cite{BB1996, KHP1999, OM2008,PP2008PRE}. Note that three different mechanisms from the DI to the DP are qualitatively
identical to those from the PCPD to the DP. In this regard, the crossover in this paper seems different
from that in Ref.~\cite{PP2006}, though the values of crossover exponents are quite close; 0.62 
in this paper and 0.57 in Ref.~\cite{PP2006}.

\begin{acknowledgments}
This work was supported by the Basic Science Research Program through the
National Research Foundation of Korea~(NRF) funded by the Ministry of
Science, ICT and Future Planning~(Grant No. 2014R1A1A2058694)
and by the Catholic University of Korea, research fund 2017.
The author furthermore thanks the Regional Computing Center of
the University of Cologne (RRZK) for providing computing time on the 
DFG-funded High Performance Computing (HPC) system CHEOPS as well as support.
The author would also like to thank Korea Institute for Advanced Study (KIAS)
for its support and hospitality during his stay there on sabbatical leave (2016-2017).
\end{acknowledgments}
\bibliography{abs.bib}
\end{document}